\def\la{\mathrel{\hbox{\rlap{\hbox{\lower4pt\hbox{$\sim$}}}\hbox{$<$}}}}
\def\ga{\mathrel{\hbox{\rlap{\hbox{\lower4pt\hbox{$\sim$}}}\hbox{$>$}}}}

\def\Teff{\ifmmode{T_{\rm eff}}\else{\hbox{$T_{\rm eff}$} }\fi}
\def\Rzero{\ifmmode{R_0}\else{\hbox{$R_0$} }\fi}
\def\kms{km s$^{-1}$}

\def\56ni{$^{56}$Ni}
\def\56co{S^{56}$Co}

\def\dm{$\Delta m_{15}$}
\def\rsi{$\mathcal{R}$(S\lowercase{i}~II)}
\def\vph{$v_{phot}$}
\def\v10{$v_{10}$(S\lowercase{i}~II)}
\def\ref{\reference{}}
\documentclass{aastex}

\received{}
\accepted{}
\journalid{}{}
\articleid{}{}
\shortauthors{Hatano et al.}
\shorttitle{}

\begin{document}
\title {On the Spectroscopic Diversity of Type Ia Supernovae}

\author {{Kazuhito Hatano\altaffilmark{1}}, {David
Branch\altaffilmark{1}}, {Eric J. Lentz\altaffilmark{1}},
{E.~Baron\altaffilmark{1}}, {Alexei~V. Filippenko\altaffilmark{2}},
and {Peter~M. Garnavich\altaffilmark{3}}}

\altaffiltext{1}{Department of Physics and Astronomy, University of
Oklahoma, Norman, Oklahoma 73019}

\altaffiltext{2}{Department of Astronomy, University of California,
Berkeley, CA 94720--3411}

\altaffiltext{3}{Department of Physics, University of Notre Dame,
Notre Dame, IN 46556}

\begin{abstract}

A comparison of the ratio of the depths of two absorption features in
the spectra of Type~Ia supernovae (SNe~Ia) near the time of maximum
brightness with the blueshift of the deep red Si~II absorption feature
10 days after maximum shows that the spectroscopic diversity of SNe~Ia
is multi--dimensional.  There is a substantial range of blueshifts at
a given value of the depth ratio.  We also find that the spectra of a
sample of SNe~Ia obtained a week before maximum brightness can be
arranged in a ``blueshift sequence'' that mimics the time evolution of
the pre--maximum--light spectra of an individual SN~Ia, the well
observed SN~1994D.  Within the context of current SN~Ia explosion
models, we suggest that some of the SNe~Ia in our sample were
delayed--detonations while others were plain deflagrations.

\end{abstract}


\section{Introduction}

Advancing our understanding of the diversity among Type~Ia supernovae
(SNe~Ia) is important for identifying the nature of the progenitor
binary systems and the explosion mechanisms, as well as for boosting
confidence in using SNe~Ia as empirical distance indicators for
cosmology.  To first order, SNe~Ia can be divided into normal events
that are highly but not perfectly homogeneous, and peculiar events
like the powerful SN~1991T (e.g., Filippenko et~al. 1992a) and the
weak SN~1991bg (e.g., Filippenko et~al. 1992b).  The next
approximation is to regard SNe~Ia as a one--parameter sequence of
events, using a photometric observable such as
\dm, the decline of the $B$ magnitude during the first 15 days after
the time of maximum brightness (Phillips 1993; Phillips et~al. 1999;
Suntzeff et~al. 1999), or using a spectroscopic observable such as
\rsi, the ratio of the depths of two absorption features near 5800 and
6100~\AA\ that ordinarily are attributed to Si~II $\lambda5972$ and
Si~II $\lambda6355$, respectively (Nugent et~al. 1995).  The \dm\ and
\rsi\ parameters are observed to be tightly correlated (Nugent et~al. 1995;
Garnavich et~al. 2000), and because they also correlate with peak
absolute magnitude they can be regarded as measures of the explosion
strength.  A one--parameter description of SNe~Ia is useful, and in
the context of Chandrasekhar--mass explosions it can be interpreted in
terms of a variation in the mass of ejected $^{56}$Ni --- although, on
the basis of explosion models the actual situation is expected to be
somewhat more complex than a one--parameter sequence (e.g.,
H\"oflich \& Khokhlov 1996; H\"oflich et~al. 1996).

Indeed, a one--parameter sequence does not completely account for the
observational diversity of SNe~Ia.  Hamuy et~al. (1996) found that
some light curves having similar values of \dm\ show significant
differences of detail, and it has been shown that a {\sl
two--parameter} luminosity correction using the $B-V$ color and \dm\
is both necessary and (so far) sufficient to standardize SNe~Ia to a
common luminosity (Tripp 1998; Tripp \& Branch 1999; Parodi
et~al. 2000).  It also has been noted that some SNe~Ia with
normal--looking spectra (i.e., having only lines of the usual ions)
have exceptionally high blueshifts of their absorption features
(Branch 1987), and Wells et~al. (1994) found a lack of correlation
between the blueshift of the Si~II $\lambda6355$ absorption near the
time of maximum brightness and the \dm\ parameter, in a small sample
of well observed SNe~Ia (see also Patat et~al. 1996). In this {\sl
Letter} we extend the discussion by showing more clearly that the
spectroscopic diversity among SNe~Ia is multi--dimensional.

\section{Data}

As a measure of the blueshift of a SN~Ia spectrum we adopt the
parameter used by Branch \& van den Bergh (1993; hereafter BvdB),
which we denote as
\v10: it is the velocity that 
corresponds to the blueshift of the flux minimum of the deep
absorption feature near 6100~\AA, attributed to Si~II $\lambda6355$,
10 days after the time of maximum brightness.  We use the data of
BvdB, as well as additional data for the following
events: SN~1984A (Barbon, Iijima, \& Rosino 1989); SN~1989B (Wells et
al. 1994); SN~1991T (Fisher et al. 1999); SNe~1992A and 1998bu (Jha et
al. 1999); SN~1994D (Patat et al. 1996; Filippenko 1997); SN~1997br
(Li et al. 1999); and SN~1999by (Bonanos et~al. 2000).  As another
spectroscopic observable we use the \rsi\ parameter.  Most of the
\rsi\ values that we use are from Garnavich et~al. (2000).  For events
for which they give several slightly different values of \rsi,
measured at several epochs, we use the median value.  For a few other
events we have estimated \rsi\ from published spectra that are
referenced by BvdB (1993), following the definition of \rsi\ by Nugent
et al. (1995).  For another nine events for which spectra suitable for
measuring \rsi\ are unavailable, we have used the observed value of
\dm\ (Phillips 1993; Schaefer 1996, 1998; Riess et ~al. 1999a; 
Suntzeff et~al. 1999) to ``predict'' the value of \rsi, with the help
of the tight relation between \rsi\ and \dm\ that is displayed by
Garnavich et~al. (2000).  The characteristic uncertainties are
500~\kms\ in \v10\ and 0.04 in
\rsi, which are not negligible but not large enough to
strongly affect our conclusions.

\section{Results}

\subsection{\rsi\ versus \v10}

The \rsi\ parameter is plotted against \v10\ in Fig.~1. If SNe~Ia
did behave as a simple one--parameter sequence, with the mass of ejected
$^{56}$Ni being the fundamental physical parameter, we would expect
\rsi\ to decrease smoothly as \v10 increases.  
The more powerful the event, the higher the velocity at the outer edge
of the core of iron--peak elements, and therefore the higher the
velocity of the silicon--rich layer.  For the sample of Fig.~1 taken
as a whole, this obviously is not the case.  Of course, it is possible
that the peculiar weak SNe~1991bg, 1999by, and 1986G, and/or the
peculiar powerful SNe~1991T and 1997br, comprise physically distinct
subgroups of SNe~Ia.  It is interesting that when we disregard
these, and restrict our attention to the spectroscopically normal
SNe~Ia, it is not even clear that there is any correlation between
\rsi\ and \v10 at all.  Normal SNe~Ia that have similar values of
\rsi\ can have substantially different values of
\v10, so a one--parameter sequence cannot completely account for the
spectroscopic diversity of normal SNe~Ia.  If we were able to use
blueshifts measured at a fixed time after explosion, rather than at a
fixed time with respect to maximum brightness, Fig.~1 would look
slightly different because of the diversity in SN~Ia rise times from
explosion to maximum brightness (Riess et~al. 1999b), but this would
not change our conclusion about the need for more than a
one--dimensional spectroscopic sequence.

\subsection{Pre--maximum spectra}

Now we turn to the diversity among pre--maximum--light spectra of
SNe~Ia.  Fig.~2 shows spectra of five events, all obtained about one
week before the time of maximum brightness, arranged in a ``blueshift
sequence.''  Some things to notice in Fig.~2 are that (1)
as is to be expected, the absorption features are broader in the
higher--blueshift events; (2) the absorption features tend to be
deeper in the higher--blueshift events (not necessarily expected); and
(3) the Ca~II absorption appears as a single feature in the
high--blueshift SNe~1984A and 1992A, while it appears split in the
lower blueshift SNe~1989B, 1994D, and 1990N.

For comparison with Fig.~2, Fig.~3 shows a time series of
pre--maximum--light spectra of the particularly well observed
SN~1994D.  The degree to which Fig.~3 resembles Fig.~2 is
intriguing.  The $-12$ day spectrum of SN~1994D in Fig.~3 looks much
like the $-7$ day spectrum of SN~1984A in Fig.~2; the SN~1994D
absorptions are at their deepest at the earliest times when the
blueshifts are the highest; and the Ca~II absorption of SN~1994D
evolves from unsplit to split.  It is almost as if normal SNe~Ia
follow a standard pattern of pre--maximum spectroscopic evolution
after all, but their spectra are not in phase with respect to the time
of maximum brightness.  (This statement should not be taken too
literally, and certainly it cannot be said of the spectroscopically
peculiar SNe~Ia; there was no pre--maximum phase at which the spectrum
of SN~1994D resembled that of SN~1991bg--like or SN~1991T--like
events.)

Fig.~4 shows some comparisons of observed pre--maximum SN~Ia spectra
with synthetic spectra that have been generated with the parameterized
spectrum--synthesis code SYNOW.  [Brief descriptions of the code can
be found in recent applications of SYNOW, e.g., Fisher et~al. (1999),
Millard et~al. (1999), Hatano et~al. (1999a); details are in
Fisher (2000).]  The top panel compares the $-6$ day spectrum of
SN~1994D with a best--fit synthetic spectrum that has a velocity at
the photosphere of
\vph$=11,000$~\kms\ and includes lines of O~I, Mg~II, Si~II, Si~III,
S~II, Ca~II, and Fe II.  The second panel compares the $-6$ day
spectrum of SN~1992A with two synthetic spectra.  In one of them
(dashed line) the SYNOW input parameters are the same as in the top
panel, except that
\vph\ has been increased from 11,000 to 13,000~\kms; this 
produces about the right absorption--feature blueshifts for SN~1992A,
but most of the synthetic absorptions are too weak.  In the other
synthetic spectrum of the middle panel (dotted line), the optical
depths of the O~I, Si~II, and S~II lines have been increased to
achieve a better fit.  Similarly, in the bottom panel the $-7$ day
spectrum of SN~1984A is compared to two synthetic spectra. In one of
them (dashed line) only \vph\ has been changed, from 13,000 to
17,000~\kms\, to get the right blueshifts.  In the other (dotted
line), the optical depths of the Mg~II, Si~II, Ca~II, and Fe~II lines
have been further increased [and a weak high--velocity Fe~II component
(cf. Hatano et~al. 1999a) has been introduced] to make the synthetic
absorptions strong enough.  Fig.~4 shows that these rather different
pre--maximum spectra of SNe~1994D, 1992A, and 1984A can be matched
well with lines of the same ions, but with different photospheric
velocities and line optical depths.  In general, the higher the
blueshift of the spectrum, the higher the line optical depths.

\section{Discussion}

Attempting an in--depth interpretation of the causes of the
spectroscopic diversity illustrated in this {\sl Letter} would be
premature, but a brief discussion is in order. As stressed above, the
main point is that spectroscopically normal SNe~Ia that have similar
values of \rsi\ (or
\dm) have a significant range in \v10, which is not what we would
expect if SNe~Ia behaved as a simple one--parameter sequence of
events.  One possible explanation is that two or more explosion
mechanisms are involved.  In this regard it is noteworthy that Lentz
et~al. (2000) find from detailed NLTE calculations that some of the
published delayed--detonation models can account for the very high
Si~II blueshift of SN~1984A.  It may be that the seven events ranging
diagonally from SN~1992A to SN~1983G in Fig.~1 are delayed detonations
having a range of strengths, while the other spectroscopically normal
events are plain deflagrations.  From this point of view it seems
useful to regard the
\rsi\ axis as a measure of the ejected mass of $^{56}$Ni and the \v10\
axis as a measure of the amount of matter ejected at high velocity;
the latter is much higher in delayed detonations than in plain
deflagrations (Lentz et~al. 2000).  It is not clear whether the weak
SN~1991bg--like events should be regarded as weak delayed detonations,
weak deflagrations, or something else.  For suggestions that the
powerful SN~1991T--like events may come from super--Chandrasekhar
white dwarf merger products, see Fisher et~al. (1999).

We should keep in mind that the distribution of events in Fig.~1 also
could be affected by asymmetries.  BvdB (1993) argued, on the basis of
a perceived connection between \v10\ and parent--galaxy type, that the
whole range of \v10\ values cannot be attributed entirely to
asymmetries, and SNe~Ia in general are observed to have low
polarization (Wang et~al. 1996), but the possibility that asymmetry
does play some role in Fig.~1 is not excluded.  The ejected matter
could be inherently asymmetric because of, e.g., a
deflagration--detonation transition that begins at a single point
rather than on a spherical shell (Livne 1999), and/or an asymmetry
could be imposed by the presence of a donor star (Marietta
et~al. 2000). 

Figs. 2, 3, and 4 raise some additional questions.  Why are the line
optical depths higher when the blueshifts are higher?  In the Sobolev
approximation, line optical depths are proportional to the time since
the explosion (owing to the decreasing velocity gradient and the
increasing size of the line resonance region), and this time
dependence is only partially offset by the decreasing density at the
photosphere.  Therefore, temperature evolution seems to be required to
account for the decreasing line strengths that we see in Fig.~3.  If
the temperature of the line--forming layers of SN~1994D increased
during the rise towards maximum brightness, [as indicated by
broad--band photometry (Richmond et al. 1995; Patat et~al. 1996;
Meikle et~al. 1996)], the Ca~II and Si~II optical depths would have
decreased as the optical depths of the weaker Si~III and Fe~III lines
increased (Hatano et~al. 1999b); this does seem to have been the case
for SN~1994D (cf. Hatano et~al. 1999a).

Temperature differences may also be partly responsible for the
blueshift sequence of Fig.~2.  The resemblance of the spectrum of
SN~1994D at $-12$ days to that of SN~1984A at $-7$ days may mean that
the high--velocity layers of SN~1984A at $-7$ days were still as cool
as those of SN~1994D at $-12$ days.  The higher densities in the
high--velocity layers of delayed detonations are likely to cause them
to be cooler than those of plain deflagrations at a fixed time after
explosion.

Our suggestion that Fig.~1 is populated by both delayed detonations
and deflagrations seems to be a reasonable working hypothesis within
the context of current SN Ia explosion models, but there are many
uncertainties associated with such models (Hillebrandt \& Niemeyer
2000).  Our understanding of the causes of the spectroscopic diversity
of SNe~Ia is still in a rudimentary state. Many more high--quality
observed spectra and much more spectroscopic analysis are needed.

We thank Darrin Casebeer, Dean Richardson, and Thomas Vaughan for
helpful discussions.  This work was supported by NSF grants
AST--9417213, 9731450, and 9987438, as well as by NASA grant
NAG5-3505.

\clearpage

\begin {references}

\ref Barbon, R., Iijima, T., \& Rosino, L. 1989, A\&A, 220, 83

\ref Bonanos, A. et al. 2000, in preparation

\ref Branch, D. 1987, ApJ, 316, L81

\ref Branch, D., \& van den Bergh, S., 1993, AJ, 105, 2231 (BvdB)

\ref Filippenko, A. V. 1997, ARAA, 35, 309

\ref Filippenko, A. V. et al. 1992a, ApJ, 384, L15

\ref Filippenko, A. V. et al. 1992b, AJ, 104, 1543 

\ref Fisher, A. 2000, PhD thesis, University of Oklahoma

\ref Fisher, A., Branch, D., Hatano, K., \& Baron, E., 1999, MNRAS, 304,
67

\ref Garnavich, P. M. et~al. 2000, in preparation

\ref Hamuy, M. et al. 1996, AJ, 112, 2438

\ref Hatano, K., Branch, D., Fisher, A., Baron, E., \& Filippenko,~A.~V.
1999a, ApJ, 525, 881

\ref Hatano, K., Branch,~D., Fisher,~A., Millard,~J., \& Baron,~E.
1999b, ApJS, 121, 233

\ref Hillebrandt, W., \& Niemeyer, J. C. 2000, ARAA, in press

\ref H\"oflich, P., \& Khokhlov, A. 1996, ApJ, 457, 500

\ref H\"oflich, P. et al. 1996, ApJ, 472, L81

\ref Jha, S. et al., 1999, ApJS, 125, 73

\ref Kirshner, R. P. et al. 1993, ApJ, 415, 589

\ref Leibundgut, B. et al., 1993, AJ, 105, 301

\ref Lentz, E., Baron, E., Branch, D., \& Hauschildt, P. 2000, 
ApJ, in press

\ref Li, W. D. et al., 1999, AJ, 117, 2709

\ref Livne, E. 1999, ApJ, 527, 97

\ref Marietta, E., Burrows, A., \& Fryxell, B. 2000, ApJS, in press

\ref Meikle, W. P. S. et al. 1996, MNRAS, 281, 263

\ref Millard, J. et~al. 1999, ApJ, 527, 746

\ref Nugent, P. E., Phillips, M. M., Baron, E., Branch,~D., \&
Hauschildt,~P. 1995, ApJ, 455, L147

\ref Parodi, B. R., Saha, A., Sandage, A., \& Tammann, G. A., 2000,
ApJ, in press

\ref Patat, F., Benetti, S., Cappellaro, E., Danziger,~I.~J., Della
Valle,~M., Mazzali,~P.~M., \& Turatto,~M. 1996, MNRAS, 278, 111

\ref Phillips, M. M., 1993, ApJ, 413, L105

\ref Phillips, M. M., Lira, P., Suntzeff,~N.~B., Schommer,~R.~A.,
Hamuy,~M., \& Maza,~J. 1999, AJ, 118, 1766 

\ref Richmond, M. W. et al. 1995, AJ, 109, 2121

\ref Riess, A. G. et al. 1999a, AJ, 117, 707

\ref Riess, A. G. et al. 1999b, AJ, 118, 2675

\ref Schaefer, B. E., 1996, ApJ, 460, L19

\ref Schaefer, B. E., 1998, ApJ, 509, 80

\ref Suntzeff, N. B. et al., 1999, AJ, 117, 1175

\ref Tripp, R. 1998, A\&A, 331, 815

\ref Tripp, R., \& Branch, D. 1999, ApJ, 525, 209

\ref Wang, L., Wheeler, J. C., Li, Z., \& Clocchiatti, A. 1996, ApJ,
467, 435

\ref Wegner, G., \& McMahan, R. K. 1987, AJ, 93, 287

\ref Wells, L. A. et al., 1994, AJ, 108, 2233

\end{references}

\clearpage

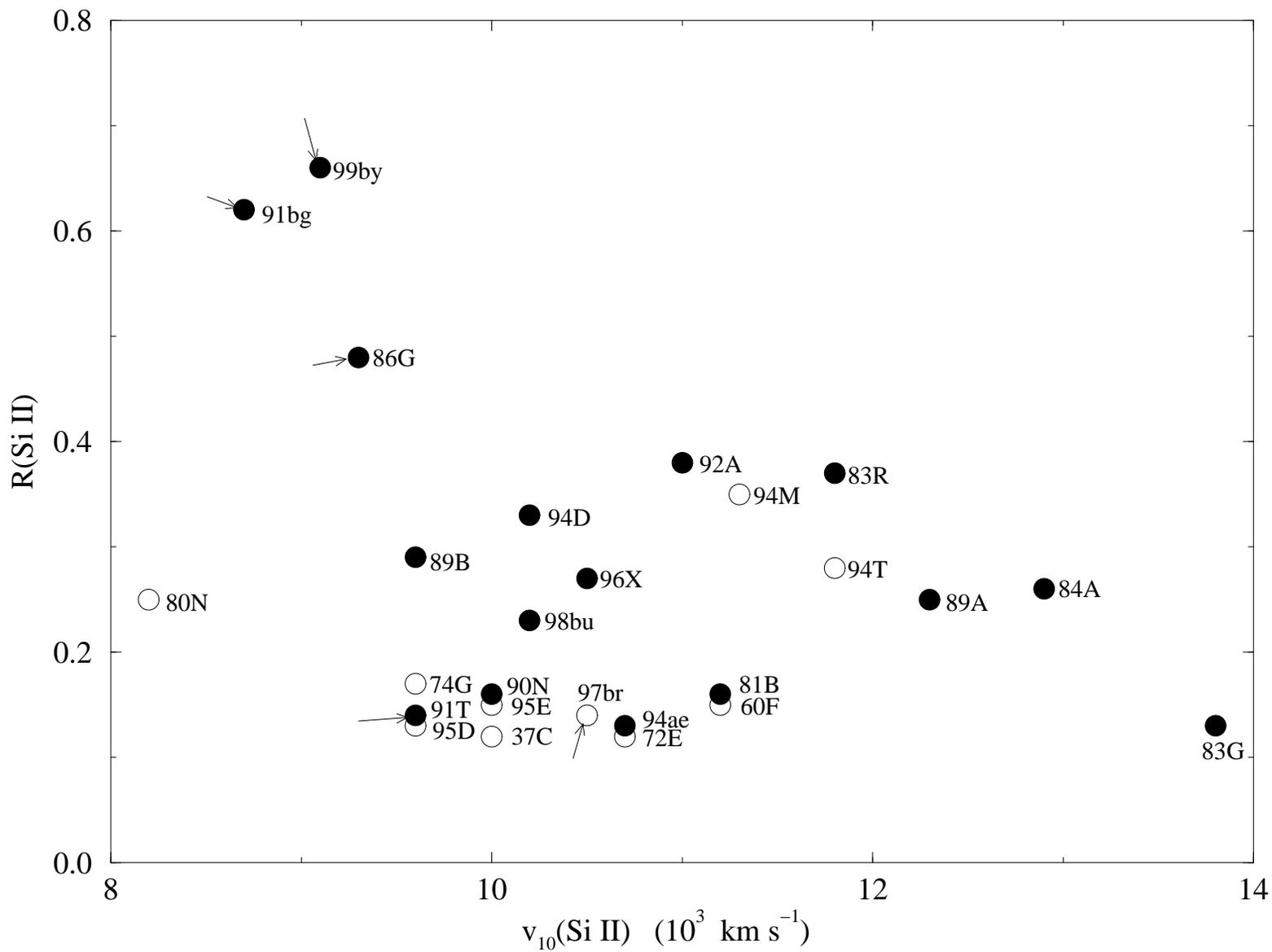
\noindent Fig. 1: The \rsi\ parameter is plotted against \v10.  
Arrows denote spectroscopically peculiar SNe~Ia.  Open symbols mean
that \rsi\ has been obtained from a relation between \dm\ and
\rsi.  

\bigskip

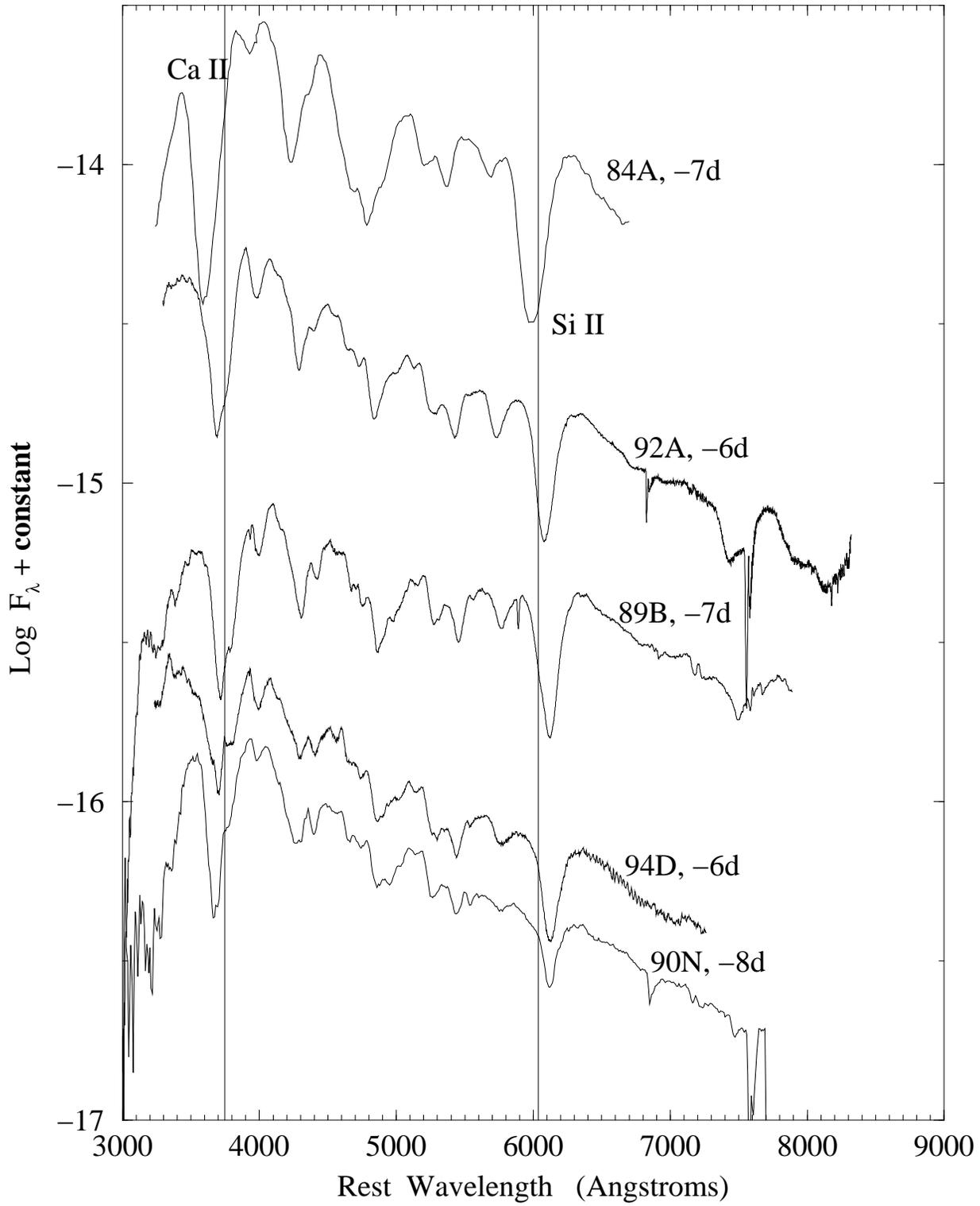
\noindent Fig. 2: Spectra of five SNe~Ia about a week before the
time of maximum light (SN~1984A: Wegner \& McMahan 1987; SN~1992A:
Kirshner et~al. 1993; SN~1989B: Wells et~al. 1994; SN~1994D:
Filippenko 1997; SN~1990N: Leibundgut et~al. 1993) are arranged in a
blueshift sequence.  The vertical lines, drawn as an aide to the eye,
are blueshifted by 15,000~\kms\ from $\lambda$6355 (Si~II) and
$\lambda$3945 (Ca~II).  The vertical displacements are arbitrary.

\bigskip

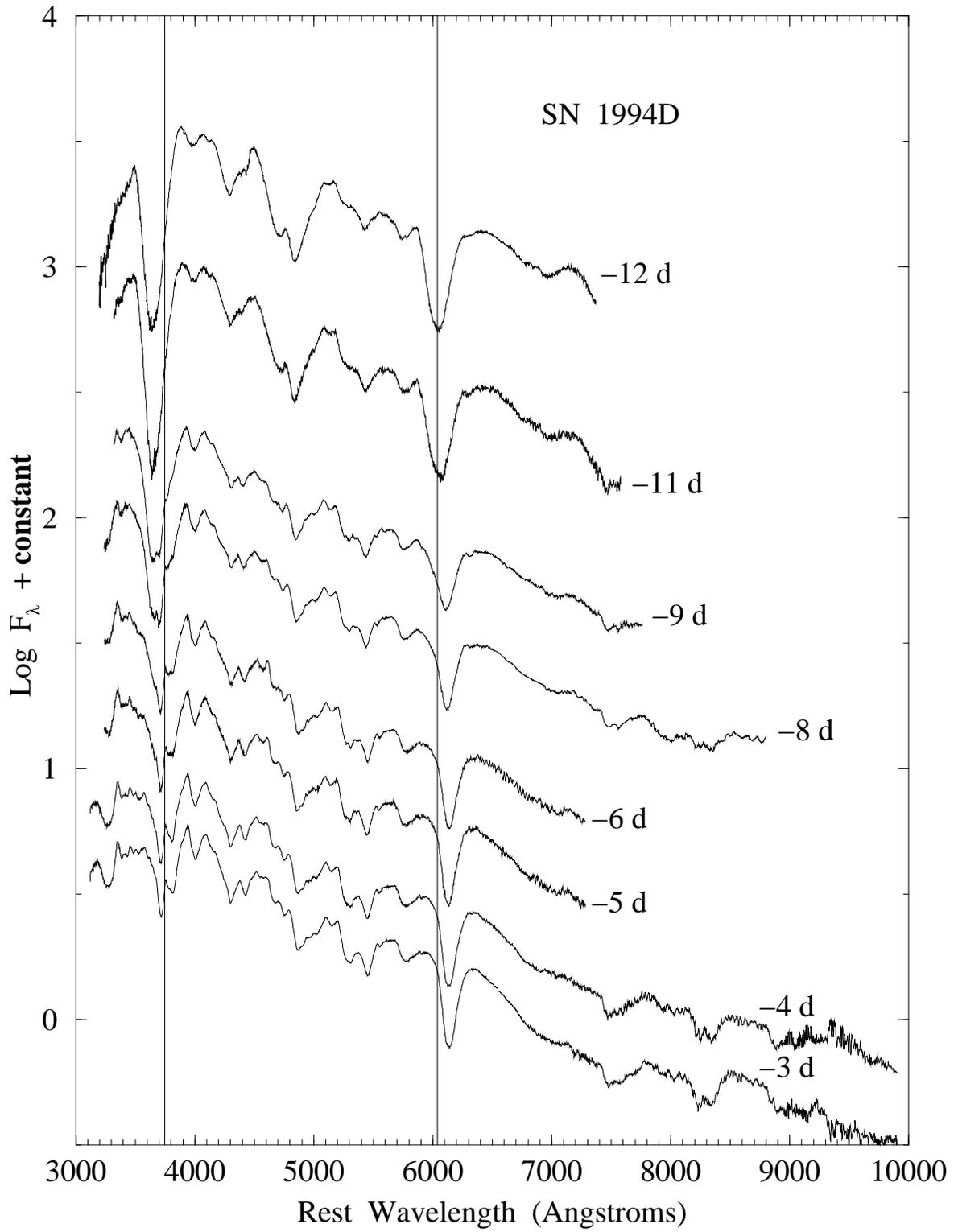
\noindent Fig.~3: A time series of pre--maximum spectra of SN~1994D
(Patat et~al. 1996; Filippenko 1997, and unpublished).  The vertical
lines are the same as in Fig.~2. 

\bigskip

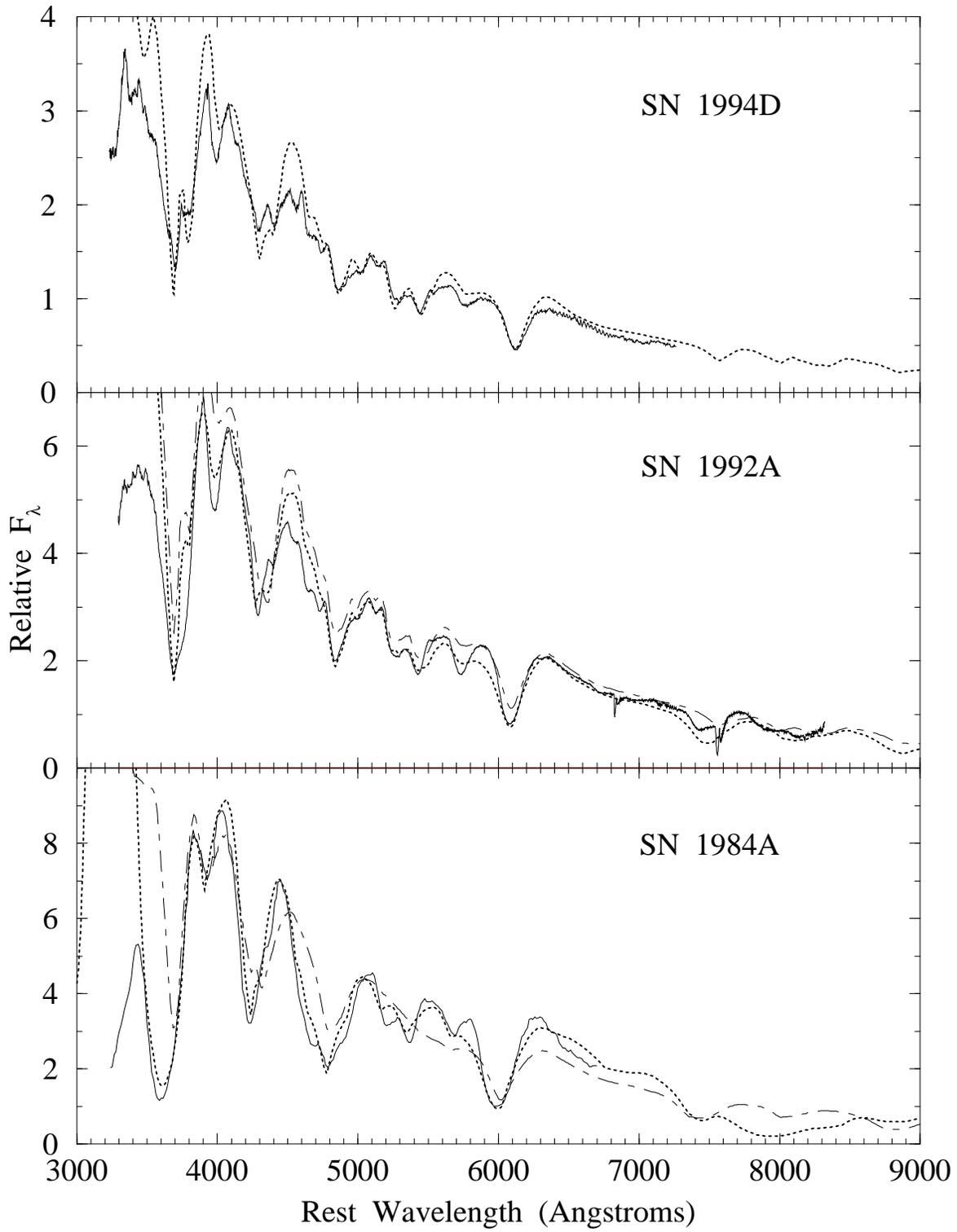
\noindent Fig.~4: Comparisons of observed spectra (solid curves) 
of SN~1994D at $-6$ days (top panel), SN~1992A at $-6$ days (middle
panel), and SN~1984A at $-7$ days (bottom panel) with SYNOW synthetic
spectra.  See the text for descriptions of the synthetic spectra.

 \end{document}